\documentclass[conference]{IEEEtran}
\IEEEoverridecommandlockouts
\usepackage{cite}
\usepackage{amsmath,amssymb,amsfonts}
\usepackage{algorithmic}
\usepackage{graphicx}
\usepackage{textcomp}
\usepackage{xcolor}
\usepackage{amsbsy}
\usepackage{amsmath,amsfonts}
\usepackage{algorithmic}
\usepackage{graphicx}
\usepackage{textcomp}
\usepackage{xspace}
\usepackage{comment}
\usepackage{enumitem}
\usepackage{mdframed}
\usepackage{url}
\usepackage{todonotes}
\usepackage{color, colortbl}
\usepackage{listings}
\usepackage{multirow}
\usepackage[hidelinks]{hyperref}
\usepackage[newfloat,frozencache,cachedir=.]{minted}

\def\BibTeX{{\rm B\kern-.05em{\sc i\kern-.025em b}\kern-.08em
    T\kern-.1667em\lower.7ex\hbox{E}\kern-.125emX}}
\begin{document}

\makeatletter
\newcommand{\linebreakand}{%
  \end{@IEEEauthorhalign}
  \hfill\mbox{}\par
  \mbox{}\hfill\begin{@IEEEauthorhalign}
}
\makeatother

\title{Testing the Unknown: 
A Framework for OpenMP Testing via Random Program Generation\thanks{This 
work was performed under
the	auspices of the	U.S. Department of Energy by Lawrence Livermore	
National Laboratory under Contract DE-AC52-07NA27344 (LLNL-CONF-868411).}
}


\author{
\IEEEauthorblockN{Ignacio Laguna}
\IEEEauthorblockA{
\textit{Center for Applied Scientific Computing} \\
Lawrence Livermore National Laboratory \\
ilaguna@llnl.gov
}
\and
\IEEEauthorblockN{Patrick Chapman}
\IEEEauthorblockA{
\textit{Department of Computer Science} \\
University of California, Davis \\
pchapman@ucdavis.edu
}
\and
\IEEEauthorblockN{Konstantinos Parasyris}
\IEEEauthorblockA{
\textit{Center for Applied Scientific Computing} \\
Lawrence Livermore National Laboratory \\
parasyris1@llnl.gov
}
\linebreakand 
\IEEEauthorblockN{Giorgis Georgakoudis}
\IEEEauthorblockA{
\textit{Center for Applied Scientific Computing} \\
Lawrence Livermore National Laboratory \\
georgakoudis1@llnl.gov
}
\and
\IEEEauthorblockN{Cindy Rubio-González}
\IEEEauthorblockA{
\textit{Department of Computer Science} \\
University of California, Davis \\
crubio@ucdavis.edu
}
}

\maketitle


\definecolor{dkgreen}{rgb}{0,0.6,0}
\definecolor{darkred}{rgb}{0.3,0.1,0.1}
\definecolor{gray}{rgb}{0.5,0.5,0.5}
\definecolor{mauve}{rgb}{0.58,0,0.82}
\definecolor{light-gray}{gray}{0.9}
\definecolor{blue}{rgb}{0,0,0.75}

\lstset{ %
  language=C,
  basicstyle=\scriptsize\ttfamily,
  numbers=left,
  numberstyle=\scriptsize\color{gray},  
  numbersep=5pt,                  
  backgroundcolor=\color{white},
  showspaces=false,               
  showstringspaces=false,         
  showtabs=false,
  frame=single,                   
  rulecolor=\color{black},
  tabsize=2,                      
  captionpos=b,                   
  breaklines=true,                
  breakatwhitespace=false,
  keywordstyle=\color{blue},          
  commentstyle=\color{dkgreen},       
  stringstyle=\color{mauve},         
  escapeinside={\%*}{*)},
  xleftmargin=4.0ex,
  morekeywords={for,each,between,can,reach,in,is,Sort,Print,From}
}

\providecommand{\en}[1]{\ensuremath{\text{E{#1}}  }}

\newcommand{\subheader}[1]{\noindent \textbf{#1}}

\newcommand{\squeezeup}{}

\hyphenation{allo-ca-tion scale-de-pen-dent}

\renewcommand{\ttdefault}{cmtt}

\newlist{rqs}{enumerate}{1}
\setlist[rqs,1]{label={\bfseries Q\arabic*},align=left,wide, labelwidth=!, 
labelindent=0pt}

\newcommand{\note}[1]{{\color{red} NOTE: #1}}

\pagestyle{plain}

\begin{abstract}
We present a randomized differential testing approach to test OpenMP 
implementations. In contrast to previous work that manually creates
dozens of verification and validation tests, our approach is able to randomly 
generate thousands of tests, exposing OpenMP implementations to a wide range 
of program behaviors.
We represent the space of possible random OpenMP tests using a grammar 
and implement our method as an extension of the \texttt{Varity} 
program generator.
By generating 1,800 OpenMP tests, we find various performance 
anomalies and correctness issues when we apply them to 
three OpenMP implementations: GCC, Clang, and Intel.
We also present several case studies that analyze the anomalies and give
more details about the classes of tests that our approach creates.
\end{abstract}


\section{Introduction}

While OpenMP is widely used, it continues to be
challenging to test OpenMP implementations. 
There are several OpenMP implementations available for C/C++ 
and Fortran---the OpenMP website lists at least 19 compilers
from various vendors and open-source community that implement 
the OpenMP API\footnote{https://www.openmp.org/resources/openmp-compilers-tools/}.
Such a wide range of interpretations of the API
can lead to different implementations of OpenMP features, which makes testing very difficult. 
Users of HPC systems are usually provided with
several OpenMP implementations, which are composed of various
components, such as a compiler and a runtime system. When an HPC system
is deployed or new versions of OpenMP are installed in the system, it is
crucial to test the implementations and identify possible performance or
correctness bugs.

Previous work on testing OpenMP implementations has been 
based on manually creating 
\textit{benchmark programs}, or tests, for verification and validation (V\&V)
of the OpenMP features available in the implementations~\cite{diaz2018openmp,muller2003openmp,wang2012openmp}. 
Such an approach has been helpful for V\&V and in identifying common bugs.
Those studies target tests for particular versions of the OpenMP API; thus, they 
can be useful in testing specific features of the OpenMP standard.
While these methods have been useful, they are limited by 
the classes of programs that the
benchmark tests encode, the different behaviors that such tests expose,
and the inputs used in the tests.

Random testing has been used as a black-box testing approach,
in which tests are generated randomly~\cite{hamlet1994random,arcuri2011random}
in contrast to the V\&V approach that creates cherry-picked tests manually. 
Randomized differential testing uses the idea that if one has multiple
implementations of the same program, all implementations
must produce the same result from the valid input.
When one implementation produces different outputs relative to the rest, 
that implementation must be faulty, or it exhibits an anomaly that must 
be further analyzed. Random testing has not been 
applied in the context of OpenMP---we seek to leverage this idea in this paper.

\textbf{Contributions.}
We present an approach to test OpenMP implementations by a 
combination of random program generation and differential testing.
Our approach generates thousands of random OpenMP tests, each test being 
composed of an OpenMP program and an input. Tests are compiled
by the different OpenMP compilers available in an HPC system. The tests
are run and evaluated for performance and correctness. 

Using randomized differential testing, we detect performance and correctness bugs 
by providing the same input to the same program compiled 
and run by different OpenMP implementations, and observing differences 
in their execution. When one execution behavior is significantly different 
from the rest, we call it an \textit{outlier}.
In contrast to previous work that manually creates dozens of specific tests,
our approach is able to randomly generate thousands of tests, exposing
OpenMP implementations to a wide variety of program behaviors.
This approach can be beneficial in uncovering performance and correctness
bugs.

\begin{figure*}[th!]
\centering
\includegraphics[width=7in]{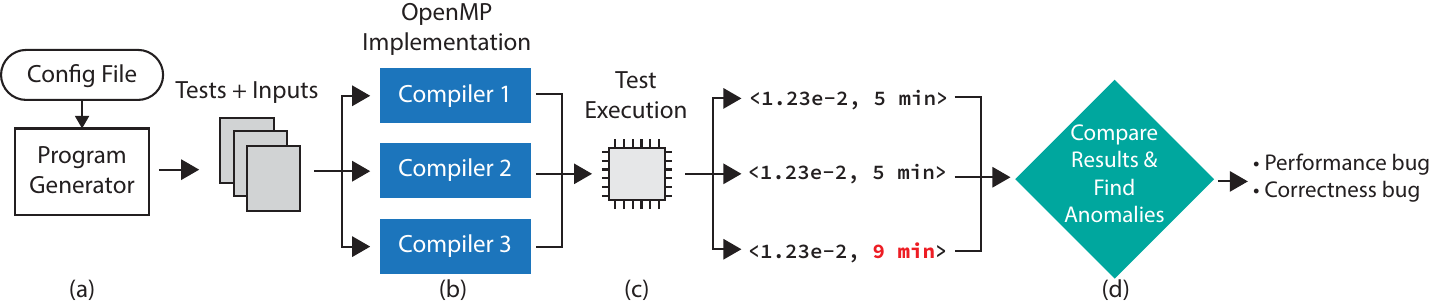}
\caption{Overview of our approach. Step (a) generates program tests and inputs; in step (b) we compile the test with multiple
OpenMP compilers---the goal is to find a bug in any of the OpenMP implementations; (in step (c) we run the tests and gather
numerical results and execution time; in step (d) we compare the numerical results and executions times and identify
bugs by finding outliers. The figure highlights am anomaly in the test produced by the OpenMP implementation 3, in which the execution time is significantly different from the execution times of the other tests (generated by the other OpenMP implementations).}
\label{fig:overview}
\end{figure*}

Our contributions are:
\begin{enumerate}
    \item A method to generate random OpenMP programs for differential
    testing. Our method generates programs with commonly used OpenMP 
    directives, such as parallel regions, for-loop regions, reductions, 
    and critical sections. The generated programs perform complex numerical
    computations, mimicking computations in scientific codes.
    
    \item An implementation of the approach in the \texttt{Varity}~\cite{varity}
    framework for floating-point random program generation. Our framework includes the 
    generation of random floating-point inputs (inherited from
    \texttt{Varity}), time execution profiling, and correctness checking
    for various categories of correctness bugs (\textit{crashes} and 
    \textit{hangs}).

    \item An evaluation of the approach on three OpenMP implementations
    (GCC, Clang, and Intel) in an HPC cluster.
    Our evaluation generates and evaluates more than 1,800
    randomly generated tests, and was able to identify various tests
    on which the OpenMP implementations exhibit performance and 
    correctness issues. We found several outlier cases where the 
    binaries produced by an implementation are either significantly slow
    or significantly fast relative to the others, as well as
    cases that induce correctness issues in the OpenMP implementations.
    We present various case studies with details of such cases.
\end{enumerate}
\section{Overview}
In this section, we present a high-level view of our approach
and an example of a generated test that exhibits a performance issue
in an OpenMP implementation.

\subsection{Workflow Overview}
Our approach's workflow is shown in Figure~\ref{fig:overview}.
The generator first uses a configuration file to obtain the parameters
to use in the program generation phase (step (a)). The parameters include
the compilers to use, optimization levels, the directories to save the tests,
and parameters related to the complexity of the random programs (see 
Section~\ref{sec:varity} for more details). When the programs and floating-point
inputs are randomly generated, they are compiled by the available OpenMP
compilers in the cluster (step (b)). 

There is a driver that then runs
all the binaries with their corresponding inputs in the systems (step (c)).
The driver checks the outputs of the tests and whether there is a correctness
issue with any test (e.g., the program abruptly terminates).
Finally, we compare the output and execution times of the tests (step (d))---if
for a given test, the binary for a specific implementation shows a behavior
different from the rest, we flag it as an \textit{outlier}. For example,
in Figure~\ref{fig:overview}, the execution time for the binary produced 
by the compiler 3 was 9 minutes, while the other binaries took 5 minutes,
for the same input and test. The binary for compiler 3 is flagged as an outlier
in this case. This test can now be investigated in more detail to determine
if the OpenMP implementation has a bug that is exposed by this test.

\begin{listing}[t]
\begin{minted}[xleftmargin=1em, breaklines, linenos, frame=lines,numbersep=2pt,fontsize=\footnotesize]{cpp}
void compute(double* comp, int var_1, ...) {
  ... 
  for (int i=0; i < var_1; ++i) {
    ...
    comp[i % 1000] += var_2[id] - -1.0 * var_3 * var_4;
    ...
    #pragma omp parallel default(shared) private(var_1, ...) firstprivate(var_2, ...) num_threads(36)
    {
    var_1 = 0;
    #pragma omp for
    for (int i=0; i < var_6; ++i) {
      ...
      comp[id] += ...;
    }
    }
...
\end{minted}
\caption{Random test that induces a performance anomaly in Clang.}
\label{fig:example_test}
\end{listing}

\subsection{Example of Generated Test}
As a ``teaser'' for the reader, we present an example of 
the tests that our approach can generate.
Section~\ref{sec:examples} presents more examples at a higher
level of detail. Listing~\ref{fig:example_test} shows the
test example. When the test was compiled with three OpenMP
implementations---Clang, GCC, and Intel---in an \texttt{x86}
system, the execution time of the Clang binary was $10\times$
higher than the execution time for the binaries
produced with the other OpenMP implementations (Intel and GCC).

As Listing~\ref{fig:example_test} shows, the random test has a
parallel region inside a loop, which stresses the capabilities of
the OpenMP runtime system for invoking and launching worker tasks.
This test exposes a deficiency in the Clang implementation
for managing tasks resources relative to the efficiency of the
other implementations. We provide more details of this example
in Section~\ref{sec:clang_slow}. Note that the test along with 
the particular input that generates this behavior is found by
our approach and provided to the users for further investigation.
\section{Code Generation Approach}
In this section, we present our approach for code generation, 
which is implemented in the 
\texttt{Varity} framework. We first present an overview of \texttt{Varity}
and then we present our approach to support OpenMP parallel
programs random generation using \texttt{Varity}.

\subsection{Varity}
\label{sec:varity}
We based our framework on the \texttt{Varity}~\cite{varity}
random program generator. \texttt{Varity} generates random
programs that expose a wide range of floating-point arithmetic
operations, and other structures encountered in scientific codes,
such as, \texttt{for} loops, and \texttt{if} conditions.
\texttt{Varity} also generates random floating-point
inputs for the programs.
\texttt{Varity} was originally designed for serial programs.
In this work, we extend it to support parallel OpenMP programs
and perform performance testing---originally, the framework
only included support for testing numerical correctness issues.

\lstdefinestyle{myCustomStyle}{
  language=C,
  basicstyle=\footnotesize\ttfamily,
  linewidth=.75\textwidth
}
\lstset{style=myCustomStyle}
\begin{listing*}[t]
\begin{minted}[xleftmargin=1em, linenos, frame=lines,numbersep=2pt,fontsize=\footnotesize]{cpp}
/** Function-level rules **/
<function> ::= "void" "compute" "(" <param-list> ")" "{" <block> "}"
<param-list> ::= <param-declaration> | <param-list> "," <param-declaration>
<param-declaration> ::= "int" <id> | <fp-type> <id> | <fp-type> "*" <id>

/** Expression- and term-level rules **/
<assignment> ::= "comp" <assign-op> <expression> ";" | <fp-type> <id> <assign-op> <expression> ";"

<expression> ::= <term> | "(" <expression> ")"  | <expression> <op> <expression>
<term> = <identifier> | <fp-numeral>

/** Block-level rules **/
<block> ::= {<assignment>}+ | <if-block> <block> | <for-loop-block> <block> | <openmp-block>

/** OpenMP-block-level rules **/
<openmp-head> ::= "#pragma omp parallel default(shared) private("<private-vars> ")" 
                  " firstprivate("<first-private-vars>")" {" reduction(" <reduction-op> ": comp)"}?
<openmp-block> ::= <openmp-head> "\n{" {<assignment>}+ <for-loop-block> "}"
<openmp-critical> ::= "#pragma omp critical {\n" <block> "}"

/** If-block-level rules **/
<if-block> ::= "if" "(" <bool-expression> ")"  "{" <block> "}"

/** For-loop-level rules **/
<for-loop-head> ::= "#pragma omp for \n for" | "for"
<for-loop-block> ::= <for-loop-head> "(" <loop-header> ")" "{" {<block>|<openmp-critical>}+ "}"
<loop-header> ::= "int" <id> ";" <id> "<" <int-numeral> ";" "++" <id> 

/** Bool-expresion-level rules **/
<bool-expression> ::= <id> <bool-op> <expression>
\end{minted}
\caption{High-level specification of the grammar for the random test programs. \texttt{<fp-type>} 
supports \{\texttt{float}, 
\texttt{double}\}, 
\texttt{<assignment-op>} supports \{\texttt{=}, \texttt{+=}, \texttt{-=}, \texttt{*=}, 
\texttt{/=}\},
\texttt{<op>} supports \{\texttt{+}, \texttt{-}, \texttt{*}, \texttt{/}\}, and 
\texttt{<bool-op>} supports \{\texttt{<}, \texttt{>}, \texttt{==}, \texttt{!=}, 
\texttt{>=}, \texttt{<=}\}. \texttt{<fp-numeral>} is a constant, e.g., $1.23\text{e+}4$.
\texttt{<reduction-op>} supports \{\texttt{+}, \texttt{*}\}.
}
\label{fig:grammar}
\end{listing*}

\begin{figure}[h]
    \centering
    \includegraphics[width=3in]{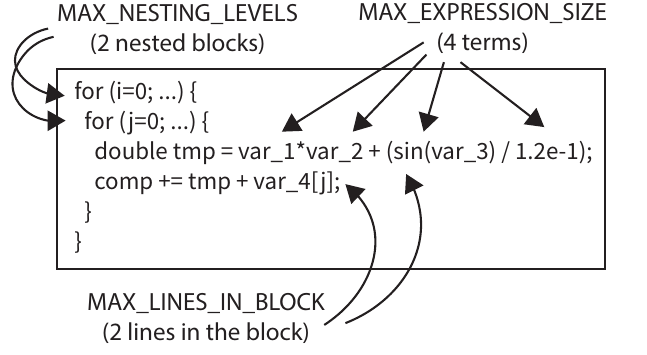}
    \caption{Example of how parameters control the code generation.}
    \label{fig:parameters}
\end{figure}

\subheader{Grammar.}
While ideally, one would want to explore the entire space of 
all possible OpenMP programs, this is not practical. Therefore, we restrict our search to a subset of 
programs: programs written in the \texttt{C++} 
language with a well-defined structure. To 
formally define this structure, we use a grammar.
Listing~\ref{fig:grammar} presents a high-level overview of the
grammar, including the extensions to support OpenMP parallel 
regions.

\texttt{Varity}'s grammar already considers the most important 
aspects of HPC programs and uses the characteristics of programs 
that could (most likely) affect how floating-point code is 
generated and executed. The grammar allows us to generate 
programs with the following characteristics:
\begin{itemize}[leftmargin=*]
\item \textbf{Different Floating-Point Types:} 
we can generate variables using single and double floating-point precision 
(i.e., \texttt{float} and \texttt{double}).

\item \textbf{Arithmetic Expressions:} arithmetic expressions can use any 
operator in \{\texttt{+}, \texttt{-}, \texttt{*}, \texttt{/}\}, 
can use parenthesis ``()'', and can use functions from the \texttt{C math} library. 
The grammar also allows boolean 
expressions.

\item \textbf{Loops:} loops constitute the main 
building block of HPC programs; the grammar allows the generation of \texttt{for} 
loops with multiple levels of nesting. We can generate loop sets $L_1 > L_2 > L_3 > \ldots > L_N$, where 
$L_1$ encloses $L_2$, $L_2$ encloses $L_3$, 
and so on up to $L_N$, where $N$ is defined by the user.

\item \textbf{Conditions:} the grammar supports \texttt{if} conditions, 
which can be true or false based on a boolean expression.

\item \textbf{Variables:} programs can contain temporary 
floating-point variables. 
Variables can store arrays or single values.
\end{itemize}

\begin{figure*}[t]
\centering
\includegraphics[width=7in]{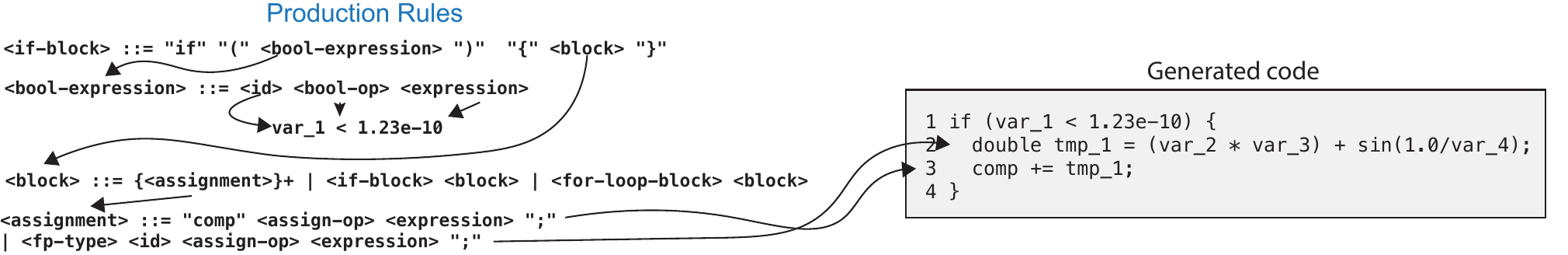}
\caption{Example of production rules representing the code generated with an if-condition block, and two assignments.
One of the assignments (line 2) has a arithmetic expression with several terms.}
\label{fig:rules_1}
\end{figure*}

\subheader{Example of Code Generation.}
Figure~\ref{fig:rules_1} shows an example of an \texttt{if-condition} block,
and the corresponding production rules that explain the generated code.
The \texttt{<if-block>} specifies that the block should have an ``\texttt{if}''
symbol, followed by an opening parenthesis ``\texttt{(}'', followed by a
\texttt{<bool-expression>}, a closing parenthesis ``\texttt{)}'', and finally
a ``\texttt{<block>}'' enclosed in brackets. The Figure shows how each element
of the code is specified in different grammar production rules. The code inside
the block comprises one ore more assignments (in this case), but it could also contain
another \texttt{<if-block>} or a \texttt{<for-loop-block>}.
Note that expressions can be arbitrarily large, containing long sequences of arithmetic
operations with variables, scalars and array values. We explain later in this section
how to control the size of arithmetic expressions.

\subsection{Program Output}
All operations are enclosed in a kernel function named \texttt{compute}. 
The kernel function does not return anything; instead, it computes a 
floating-point value and stores it in the 
\texttt{comp} variable. The \texttt{comp}'s value is printed 
to the standard output.
In addition to the \texttt{comp} kernel function, 
the generator produces a \texttt{main()} function and code to 
allocate and initialize arrays (if arrays are used in the test program). 
For simplicity, we do not present this in the grammar. The 
\texttt{main()} function reads the program inputs and copies them to the 
\texttt{comp} kernel function parameters before calling the kernel function.

\subsection{Program Generation and Randomness}
\label{sec:prog_generation}
We use randomness in the generation of test programs. 
We use the same approach that is used in previous work~\cite{csmith}
to construct a random program, i.e., uniform distributions are used to
choose elements of the program. The following features are chosen randomly 
in \texttt{Varity}: (1) type of arithmetic operations, (2) type 
of boolean operations, (3) size of arithmetic expressions, (4) size of 
boolean expressions, (5) size of blocks (i.e., number of statements), and 
(6) number of nesting levels of blocks.

\texttt{Varity} imposes limits on the above parameters since exploring infinite sets of them 
would be infeasible. The following parameters are used to limit the generation of 
program features (see Figure~\ref{fig:parameters}):
\begin{itemize}[leftmargin=*]
	\item \texttt{MAX\_EXPRESSION\_SIZE}: defines the maximum number of terms 
	in an expression (arithmetic or boolean).
	\item \texttt{MAX\_NESTING\_LEVELS}: defines the maximum number of nesting 
	levels of blocks (\texttt{if} condition and \texttt{for} loop blocks).
	\item \texttt{MAX\_LINES\_IN\_BLOCK}: a block can have several lines 
	containing, e.g., temporary variable definitions or assignments. This 
	defines the maximum number of lines in a block.
	\item \texttt{ARRAY\_SIZE}: maximum number of elements in arrays.
	\item \texttt{MAX\_SAME\_LEVEL\_BLOCKS}: in addition to assignments (or other 
	expressions), a block can have other blocks. This defines the maximum number 
	of blocks at the same nesting level inside a block.
	\item \texttt{MATH\_FUNC\_ALLOWED}: defines whether or not to use functions 
	from \texttt{math.h} in arithmetic expressions.
	\item \texttt{INPUT\_SAMPLES\_PER\_RUN}: defines the number of distinct 
	sample inputs used per program test.
\end{itemize}

\subsection{Input Generation}
\label{sec:input_generation}
Floating-point inputs are generated via an input generation 
module. This module can generate five kinds of floating-point numbers: 
\textit{normal} numbers, \textit{subnormal} numbers, \textit{almost infinity} 
numbers, \textit{almost subnormal} numbers, and 
zero (positive and negative). The normal, subnormal, 
and zero numbers correspond to those defined in the IEEE 754-2008 Standard. 
Almost infinity and almost subnormal numbers are extreme cases,
which are not defined in the Standard. We 
define an almost infinity number as a number, 
which is close to infinity 
(\texttt{+INF} or \texttt{-INF}), but that it still a normal number. 
We define an almost subnormal number as a number that is close 
to being a subnormal number, but that it is still a normal number.

\begin{figure*}[th!]
\centering
\includegraphics[width=7in]{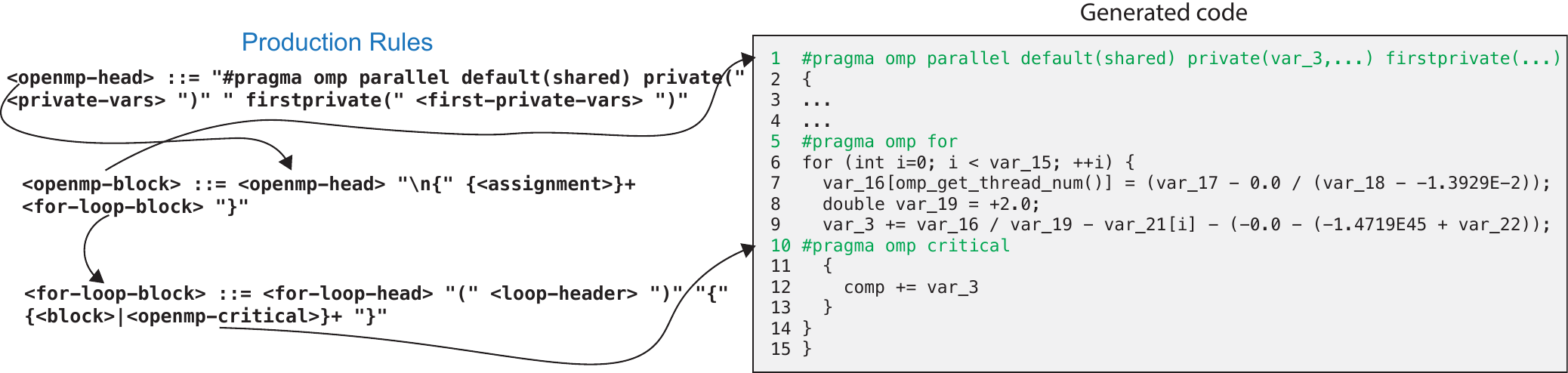}
\caption{Example of production rules used in the generation of an OpenMP block in the test.}
\label{fig:openmp_rules_1}
\end{figure*}

\subsection{OpenMP Parallel Regions}
We have extended \texttt{Varity} to support a number of OpenMP directives
that are commonly used in OpenMP programs,
such as parallel regions, parallel for loops, reductions, and critical regions.
In the following sections, we explain the OpenMP functionality we support
and the grammar rules that explain OpenMP code that is possible to generate.
In the following description, we will use as an example the OpenMP code
shown in Figure~\ref{fig:openmp_rules_1}.

The most basic OpenMP directive supported in our approach is the
the \texttt{omp parallel} directive, which instructs the 
compiler to parallelize the chosen block of code. In this directive,
we support the data-sharing OpenMP clauses \texttt{private} (for variables private to each thread),
\texttt{firstprivate} (for variables private to each thread, initialized),
\texttt{default}, and \texttt{reduction}.


Program variables are assigned to data-sharing clauses randomly except for the \texttt{comp} variable and any parallel loop-binding variable. 
This occurs when a parallel for-loop is generated and the
variable data-sharing attributes are assigned as properties of the variables.  Whenever a variable with the \texttt{shared} property is
accessed, then that code block is marked as critical in the program. 
The \texttt{comp} variable is always a shared variable unless 
it is being used in a reduction.

In Figure~\ref{fig:grammar}, the grammar shows the rules for such directives,
for the non-terminals \texttt{<openmp-block>} and \texttt{<openmp-head>}.
When the generator module decides to generate a block of code, it can choose
among different classes of blocks (e.g., \texttt{if} block), including an
OpenMP block. Note that an OpenMP block can contain other blocks. However,
this could lead to correctness problems---data races, for example---if we are not
careful about how to generate code that is accessed by all threads.
Later in this section, we discuss how we avoid data races by controlling 
the use of certain parts of the rules.

\subsection{OpenMP Reductions}
We support having a reduction clause in the \texttt{omp parallel} directive,
to perform a reduction on one variable using a specific operator (e.g., $+$). 
The reduction clause is shown in the grammar (Figure~\ref{fig:grammar}) as
\begin{equation*}
    \texttt{{" reduction(" <reduction-op> ": comp)"}?},
\end{equation*}
at the end of the \texttt{<openmp-head>} rules. This indicates that a reduction
clause can be generated one or zero times (i.e., a region may or may not have
a reduction), and the reduction variable is always the \texttt{comp} variable.
This simplifies our approach, as it allows us to keep only one reduction
variable in the region. In the future, we will explore using multiple reduction
variables.

\subheader{OpenMP Critical Sections.}
Critical sections can be generated inside \texttt{for-loop-block} regions, 
and can contain blocks of different sizes. The production rule
\texttt{<openmp-critical>} describes this clause.


\subsection{Correctness Considerations}
The code generator follows several considerations when generating
code for OpenMP regions to avoid concurrency bugs, such as 
introducing data races:
\begin{itemize}[leftmargin=*]
\item For write accesses to shared arrays,
the generator may generate variable assignments using this form:
    \texttt{var\_1[thread\_id] = var\_2 +} \ldots,
where \texttt{thread\_id} is obtained calling
\texttt{omp\_get\_thread\_num()} routine, which returns the thread
number, within the current team, of the calling thread.

\item The \texttt{comp} variable can be written inside the region,
as long as it is part of a reduction, in which case a private copy
of the variable is maintained in the local thread.

\item Concurrent accesses are enclosed in a critical section, when 
those accesses are not protected with any of the above considerations.
This prevents multiple threads from accessing the critical section 
code at the same time, thus only one active thread can update the 
data.
\end{itemize}

\subsection{Time Measurements}
We use the \texttt{std::chrono} C++ time library to obtain
the execution time of programs, using a granularity
of \texttt{chrono::microseconds}. The main computation of a test,
is contained in the \texttt{compute} function. We add timers in the
beginning and end of this function to compute the execution time.
We measure a single execution time per experiment---the time is 
printed as part of the output of the test.
\section{Outlier Detection Approach}
In this section, we present our method to detect performance or correctness issues
in the OpenMP implementations via differential testing and outlier detection.

\subsection{Assumptions and Definitions}

We consider an HPC system that has available to users several
OpenMP implementations, possibly developed by different vendors or organizations,
all following the OpenMP Language specification.
For simplicity, we assume that a system has three implementations available,
which we denote by $\{\textit{OpenMP}_1, \textit{OpenMP}_2, \textit{OpenMP}_3\}$;
however, our methodology can be applied to any number of OpenMP implementations.
We could assume, for example, that the three implementations available are the Intel, GNU GCC, and Clang
implementations. We also assume that an implementation $\textit{OpenMP}_i$ has associated:
compiler $\textit{Comp}_i$ and a runtime system $\textit{Run}_i$.

\subheader{Compiled Program.}
Given a test program $P$ generated by the code generator,
we assume that when the compiler $\textit{Comp}_i$ compiles $P$,
it produces the binary $P_i$. Therefore, for the assumed system,
we end up with three compiled binaries $P_1, P_2$, and $P_3$.
These binaries are executed with an input $I$ that is also
generated.

\subheader{Execution Times.}
When we execute a binary $P_i$, the execution time is denoted
by $r_i$ (the ``r'' indicating run time).

\subsection{Performance Outlier Detection}
We detect performance issues or bugs via outlier detection and by
comparing the different execution times, $r_i$,
for a given program and input. The intuition behind this is that,
if there is no bug or performance issue with the OpenMP implementations,
the run times $r_i$ should be all the same or \textit{comparable}.
If there is one run time that is \textit{significantly different} from the rest,
this is an indication that there could a performance bug in the OpenMP
implementation that generated and ran that program.
See Figure~\ref{fig:outliers}.

\begin{figure}[th]
\centering
\includegraphics[width=2.5in]{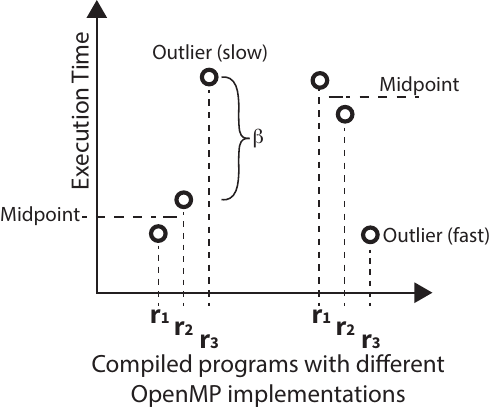}
\caption{Different classes of outliers (slow and fast) that aim to detect
with our approach.}
\label{fig:outliers}
\end{figure}

For this definition of outliers to be complete, we need to define two
important metrics: (a) when can we say that two or more execution times
are the same or \textit{comparable}? (b) when do we say that a run 
time is \textit{significantly different} from the rest? We define these terms
as follows.

\subheader{Comparable Times.}
We say that two execution times $r_i$ and $r_j$ are \textit{comparable} 
if the following is true:
\begin{equation}
    \frac{|r_i - r_j|}{min(r_i, r_j)} \leq \alpha, \ \ \ min(r_i, r_j) \neq 0.
\end{equation}
We denote two comparable execution times as $r_i \approx r_j$.
We could set $\alpha=0.2$, for example, which indicates that
the two execution times are comparable if they differ by a maximum of 20\%.
In Figure~\ref{fig:outliers}, for example, $r_1$ and $r_2$ are comparable
because they differ by a small amount relative to run time $r_3$.
When two or more execution times are comparable, we say that there is 
a \textit{midpoint} between them, denoted by $M_{i,j}$. The midpoint is the average of the comparable execution times.

\subheader{Fast and slow outliers.}
We define two classes of outliers: \textit{slow} and \textit{fast}.
We define two classes of outliers: \textit{slow} and \textit{fast}.
Suppose we have three execution times, $i$, $j$, and $k$,
and $j$, and $k$ are comparable execution times.
We say $i$ is a slow outlier if it is much higher than the
midpoint of the comparable execution times, $j$, and $k$.
Thus, the 
execution time $r_k$ is a slow outlier if the following are true:
$r_i \approx r_j$, and,
\begin{equation}
    \frac{r_k}{M_{i,j}} \geq \beta, \ \ \ M_{i,j} \neq 0.
\end{equation}
Setting $\beta=1.5$, for example, indicates that the slow
outlier is $1.5\times$ slower than the other execution times.
Similarly, we define fast outliers to indicate execution times
that are much faster than the rest.
Figure~\ref{fig:outliers} shows the two classes of outliers.

\subsection{Correctness Outliers}
Each test run is expected to produce a numerical answer, along 
with its execution time. However, the test may also stop before
finishing or may not provide an answer in the expected time, 
which could indicate that the corresponding OpenMP
implementation introduced a correctness bug. When that occurs
for a binary $P_i$, we label the execution of the binary as follows:
\begin{itemize}
\item \texttt{CRASH}: the program stopped running before providing
the final output, e.g., due to a segmentation fault, or because it 
was killed by the system;
\item \texttt{HANG}: the program took much more time to finish (relative
to the other tests), and we stopped it by sending a \texttt{SIGINT} signal.
This could occur due to various reasons, e.g., a deadlock, poor resource 
management inside the OpenMP runtime system, or others.
\end{itemize}
Note that correctness outliers are not considered to be performance outliers.

\subheader{Detection.}
If the execution for a binary $P_i$ suffered from any of the 
above cases, we denote it as $P_{i}^{\texttt{CRASH}}$ 
or $P_{i}^{\texttt{HANG}}$. If the execution terminated correctly,
we denote it as $P_{i}^{\texttt{OK}}$.
We detect correctness
outliers by checking if one execution, out of a group of executions,
exhibits either a \texttt{CRASH} or a \texttt{HANG}, while the others
did not. This could indicate that the OpenMP implementation 
has a correctness bug exposed by the test.

For example, suppose that the result of three executions are
$P_{1}^{\texttt{OK}}$, $P_{2}^{\texttt{CRASH}}$, and $P_{3}^{\texttt{OK}}$.
This indicates that both $P_1$ and $P_3$ terminated correctly, but
$P_2$ crashed, possibly indicating that we have exposed (or activated) 
a correctness bug in the the OpenMP implementation 
$\textit{OpenMP}_2$, which is the one that produced $P_2$.

\subsection{Implementation}
Our framework, including the extensions to support OpenMP code generations,
is implemented in Python 3.12. \texttt{Varity} is also originally 
implemented in Python.
We use the Python subprocess module to spawn new processes, connect to 
their input/output/error pipes and detect the test outputs,
or whether the tests suffered from a crash or hang.

\subsection{Limitations}
Our work has some limitations.
First, while the generator considers several scenarios and constraints
to generate correct OpenMP programs, we found that in some cases it can
generate data races, where the \texttt{comp} variable is written and read
by multiple threads without synchronization. We mitigated this by
manually filtering out data race cases in the evaluation.
We have identified the cause of this in \texttt{Varity}---in future work,
we will release a version that produces data-race-free programs 
100\% of the time.
Second, while the OpenMP directives we explore are widely used,
we only explore a subset of the directives from language specification---considering
more directives could lead to finding more issues.


\section{Evaluation}

In this section, we evaluate our approach on three OpenMP implementations
and summarize the results.
We designed the evaluation to answer the following research questions:

\begin{rqs}
\item Is our approach effective at finding slow/fast outliers, 
and correctness outliers in different OpenMP implementations?
\item Can the program tests associated with outliers point to
possible bugs in different OpenMP implementations?
\end{rqs}

\subsection{Evaluation System and OpenMP Implementations}
We perform all experiments in a cluster system with 2,988 nodes, where
each node has 2 18-core Intel Xeon E5-2695 processors 
(2.1 GHz) and 128 GiB of memory.
We used Python 3.12.4, and OS TOSS 4.

We use three OpenMP implementations: Intel oneAPI Compiler, 
GNU GCC, and Clang/LLVM. For a fair comparison, we used
versions released on dates close to each other:
\begin{center}
\begin{tabular}{c|c|c|c} 
 \hline
 Implementation & Compiler & Version & Release \\ \hline
 Intel oneAPI & \texttt{icpx} & 2023.2.0 & 02/2023 \\ 
  LLVM/clang & \texttt{clang++} & 16.0.0 & 03/2023 \\ 
 GCC & \texttt{g++} & 13.1 & 04/2023 \\ 
 \hline
\end{tabular}
\end{center}

We used the following configuration for \texttt{Varity}:
\texttt{MAX\_EXPRESSION\_SIZE} = 5,
\texttt{MAX\_NESTING\_LEVELS} = 3,
\texttt{MAX\_LINES\_IN\_BLOCK} = 10,
\texttt{ARRAY\_SIZE} = 1000,
\texttt{MAX\_SAME\_LEVEL\_BLOCKS} = 3,
\texttt{MATH\_FUNC\_ALLOWED} = True,
\texttt{MATH\_FUNC\_PROBABILITY} = 0.01.
For the outlier analysis, we use $\alpha = 0.2$ and $\beta = 1.5$.

We use \texttt{num\_threads(32)} to set the number of
threads to 32 (the number of cores in the system) in 
all parallel regions. We do not use any clause to specify the
thread affinity policy to be used for parallel regions.


\subheader{Number of Experiments and Execution Time.}
We generate 200 program tests (source code).
For each program test, we generate 3 different numerical inputs.
All tests are compiled with \texttt{-O3} optimization level,
with different OpenMP implementations.
In total, we run $3$ (compilers) $\times$ $200$ (programs) $\times$
$3$ (inputs) = $1,800$ execution runs.
When analyzing the results, we filter out tests that take
less than $1,000$ microseconds. This produces a total of 454 tests
to analyze.

\begin{table}[th!]
\renewcommand{\arraystretch}{1.2}
    \centering
    \caption{Overview of the results using three OpenMP implementations (Clang, GCC, and Intel).}
    \begin{tabular}{cccccc}
    \hline
    & \multicolumn{4}{c}{Outliers} \\ \cline{2-5}
    & Slow & Fast & Crash & Hang \\ \hline
    Clang & 10 & -- & -- & -- \\
    GCC & 4 & 115 & 3 & --\\
    Intel & -- & 1 & -- & 1 \\ \hline
    \end{tabular}
    \label{tab:results}
\end{table}

\subsection{Results Overview}
Table~\ref{tab:results} presents an overview of the results, showing the
number of outliers, and average execution time of the generated tests.
We first explain these results at a high-level, and then provide
several cases studies that give more details about these cases and their
potential root cause.

\subheader{Fast and Slow Outliers.}
The binaries coming from the Intel oneAPI OpenMP implementation
exhibit the smallest amount of performance outliers---we did not
observe slow outlier cases for the Intel implementation. This is expected
since the testbed platform is an Intel architecture platform, and the
Intel OpenMP compilers and runtime are expected to have the best performance
in this platform and be the ``baseline'' in terms of performance.
We observe a good number of slow outliers for Clang (10) and GCC (4).
We consider a Clang slow outlier later in Case study 2.

We observe a significant number of fast outliers for GCC binaries.
We will provide more insights into some of these in the next cases studies.
A significant number of the GCC fast outliers---about half of them---can be
attributed to numerical exceptions, such as not-a-number (NaN) values, that
impact the control flow of the tests in the GCC binaries relative to the
control-flow of the other binaries (Clang and Intel).
When these exceptions affect branching, the GCC binaries end up performing
fewer computations and producing a different numerical result than the others.
For the case studies that we present later, we only consider cases
where all the binaries produce the same numerical result.

\subheader{Correctness Outliers.}
We observe only four correctness outliers---three crash outliers
$P_{2}^{\texttt{CRASH}}$
from GCC binaries, and one $P_{3}^{\texttt{HANG}}$ case from the Intel
implementation. Thus, only $0.22\%$ out of the $1,800$ runs produce correctness
outliers. This shows that current OpenMP implementations are very reliable
when we consider such correctness anomalies. We observe no correctness outliers from the 
Clang binaries. In the next sections, we
present more details of a crash and hang outlier.


\begin{mdframed}[backgroundcolor=light-gray] \textbf{Answer to Q1:}
Our approach is effective in generating fast and slow outliers, as well
as correctness outliers (crash and hang cases). 
Out of the $1,800$ test runs, $7.4\%$ were considered outliers for our
configuration of $\alpha$, $\beta$, and the \texttt{Varity} parameters.
Changes to these parameters may produce more or less outliers.
\end{mdframed}

\subsection{Case Study 1: GCC Binary is Fast}
\label{sec:examples}
We give more details about a GCC fast outlier.\footnote{
In the dataset released with this paper, this case is
in the file \texttt{quartz1247\_532344/\_tests/\_group\_7/\_test\_2.cpp}
}
We observe that the execution time for the GCC binary
is $80\%$ faster relative to the execution time of the 
other binaries. This is a clear case of a fast outlier
for a GCC binary.

To understand the differences in performance, we use the Linux \texttt{perf} tool
(also called \texttt{perf\_events}) and gather call stack traces 
and performance counters statistics. 
We compare the GCC binary with the Intel binary because Intel represents the
baseline implementation for this platform.
Figure~\ref{fig:case1} shows the 
call stack run time overhead for both binaries. Both binaries spend a considerable 
amount of time in wait function calls---the Intel binary in \texttt{\_\_kmp\_wait}
operations and the GCC binary in \texttt{do\_wait} operations.
Since these are very different implementations of OpenMP, and wait operations
may mean different things in the implementations, it is tricky to find anomalies
by simply comparing stack traces.

\begin{figure*} 
    \begin{lstlisting}[
        xleftmargin=0.1\textwidth,
        linewidth=0.9\textwidth, 
        basicstyle=\scriptsize\ttfamily,
        numbers=none,
        caption={Intel stack traces}
        ] 
 Overhead  Command  Shared Object     Symbol
   30.85%  _test_2  libiomp5.so       [.] _INTERNALf63d6d5f::__kmp_wait_template<...
   12.13%  _test_2  libiomp5.so       [.] __kmp_wait_4
    2.84%  _test_2  [unknown]         [k] 0xffffffffae760284
    2.76%  _test_2  libiomp5.so       [.] kmp_flag_native<unsigned long long, ...
    2.00%  _test_2  [unknown]         [k] 0xffffffffae760282
    1.76%  _test_2  libiomp5.so       [.] _INTERNALf63d6d5f::__kmp_hyper_barrier_gather
    1.59%  _test_2  libiomp5.so       [.] __kmp_eq_4
    1.26%  _test_2  [unknown]         [k] 0xffffffffaf20006d
    1.10%  _test_2  libiomp5.so       [.] __kmp_hardware_timestamp
    \end{lstlisting}

    \begin{lstlisting}[
        xleftmargin=0.1\textwidth,
        linewidth=0.9\textwidth, 
        basicstyle=\scriptsize\ttfamily,
        numbers=none,
        caption={GCC stack traces}
        ]
 Overhead  Command  Shared Object      Symbol
   72.53%  _test_2  libgomp.so.1.0.0   [.] do_wait
    6.55%  _test_2  libgomp.so.1.0.0   [.] do_spin
    2.06%  _test_2  libgomp.so.1.0.0   [.] gomp_mutex_lock_slow
    1.59%  _test_2  [unknown]          [k] 0xffffffffae760282
    1.29%  _test_2  [unknown]          [k] 0xffffffffae934730
    1.23%  _test_2  [unknown]          [k] 0xffffffffae760284
    0.69%  _test_2  ld-2.28.so         [.] _dl_lookup_symbol_x
    \end{lstlisting}
\caption{Call stack overhead stats for the GCC and Intel case study 1.}
\label{fig:case1}
\end{figure*}

\begin{table}[]
 \caption{Performance counter statistics for Case Study 1.}
    \centering
    \begin{tabular}{|c|c|c|}
    \hline
    Counters & Intel & GCC \\ \hline
    context-switches & 232 & 10 \\
    cpu-migrations & 96 & 0 \\
    page-faults & 627 & 226 \\
    cycles & 110,520,780 & 154,797,061 \\
    instructions & 85,366,729 & 60,084,059 \\
    branches & 20,832,349 & 20,582,275 \\
    branch-misses & 182,300 & 67,406 \\ \hline
    \end{tabular}
\label{table:case1_counters}
\end{table}

We gather performance counter statistics, which are shown in
Table~\ref{table:case1_counters}.
We observe that the Intel binary shows many more CPU migrations, context switches, 
page faults and instructions compared to the GCC binary.
Looking into the code reveals that the generated test case
contains an OpenMP critical section, inside a parallel for loop;
the critical section updates the \texttt{comp} variable.
We speculate that the OpenMP Intel implementation suffers
from poor performance, perhaps associated with thread contention
on the critical regions, for this case.
While understanding the root cause requires more analysis---and
perhaps more sophisticated tools---this shows that our framework
can identify such outlier cases and provide interesting performance
tests that uncover corner cases.

\subsection{Case Study 2: Clang Binary is Slow}
\label{sec:clang_slow}
Here we analyze a slow outlier produced by the Clang
implementation.\footnote{
Refer to file \texttt{quartz228\_342786/\_tests/\_group\_5/\_test\_10.cpp}}
In this case, the execution time of the Clang binary is $946\%$ slower
than the rest of the binaries. Again, we compare the Clang binary
with the Intel binary, since Intel represents the baseline OpenMP implementation.
We first compare stack traces overheads, which are shown in Figure~\ref{fig:case2}.
We use the \texttt{--children} option in \texttt{perf} that 
accumulates the call chain of children to parent 
entries. The children's overhead is calculated by adding all period values
of the child functions so that it can show the total overhead of
the higher level functions. Note that in this mode, the sum of all the children's 
overhead values exceeds $100\%$.  

\begin{table}[]
 \caption{Performance counter statistics for Case Study 2.}
    \centering
    \begin{tabular}{|c|c|c|}
    \hline
    Counters & Intel & Clang \\ \hline
    context-switches & 300 & 40,483 \\
    cpu-migrations & 93 & 126 \\
    page-faults & 684 & 70,990 \\
    cycles & 1,195,535,760 & 10,168,915,718 \\
    instructions & 887,175,940 & 8,212,422,901 \\
    branches & 250,167,701 & 2,163,265,059 \\
    branch-misses & 458,225 & 3,827,212 \\ \hline
    \end{tabular}
\label{table:case2_counters}
\end{table}

\begin{figure*} 
    \begin{lstlisting}[
        xleftmargin=0.05\textwidth,
        linewidth=0.98\textwidth, 
        basicstyle=\scriptsize\ttfamily,
        numbers=none,
        caption={Intel stack traces}
        ] 
 Children     Self   Command   Shared Object          Symbol
   90.28%     0.00%  _test_10  libc-2.28.so           [.] __GI___clone (inlined)
   89.31%     0.00%  _test_10  libpthread-2.28.so     [.] start_thread
   89.00%     0.00%  _test_10  libiomp5.so            [.] _INTERNAL1ebb3278::__kmp_launch_worker
   88.95%     0.21%  _test_10  libiomp5.so            [.] __kmp_launch_thread
   71.20%    62.99%  _test_10  libiomp5.so            [.] _INTERNALf63d6d5f::__kmp_wait_template<...
   70.72%     1.32%  _test_10  libiomp5.so            [.] _INTERNALf63d6d5f::__kmp_hyper_barrier_release
   69.32%     0.00%  _test_10  libiomp5.so            [.] kmp_flag_64<false, true>::wait (inlined)
   56.49%     0.12%  _test_10  libiomp5.so            [.] __kmp_invoke_task_func
   56.34%     0.05%  _test_10  libiomp5.so            [.] __kmp_invoke_microtask
   42.75%     0.58%  _test_10  libiomp5.so            [.] __kmpc_barrier
    \end{lstlisting}

    \begin{lstlisting}[
        xleftmargin=0.05\textwidth,
        linewidth=0.98\textwidth, 
        basicstyle=\scriptsize\ttfamily,
        numbers=none,
        caption={Clang stack traces}
        ]
 Children     Self   Command   Shared Object          Symbol
   93.52%     0.00%  _test_10  libc-2.28.so           [.] __GI___clone (inlined)
   93.45%     0.00%  _test_10  libpthread-2.28.so     [.] start_thread
   93.40%     0.00%  _test_10  libomp.so              [.] 0x00001555547a46c3
   92.59%     0.00%  _test_10  libomp.so              [.] 0x00001555547488bf
   92.59%     0.02%  _test_10  libomp.so              [.] __kmp_invoke_microtask
   92.57%     0.17%  _test_10  test_10                [.] .omp_outlined.
   89.29%     0.00%  _test_10  libomp.so              [.] 0x0000155554747f51
   48.78%     0.00%  _test_10  libc-2.28.so           [.] __calloc (inlined)
   46.83%     0.01%  _test_10  [unknown]              [k] 0xffffffffafc000e9
   46.74%     0.06%  _test_10  libc-2.28.so           [.] _int_malloc
   46.68%     0.22%  _test_10  libc-2.28.so           [.] sysmalloc
   46.10%     0.00%  _test_10  [unknown]              [k] 0xffffffffaf0053eb
   44.11%     0.00%  _test_10  libc-2.28.so           [.] __GI___mprotect (inlined)
   43.83%     0.00%  _test_10  [unknown]              [.] 0xffffffffaf2f890b
   43.19%     0.02%  _test_10  libomp.so              [.] __kmpc_barrier
    \end{lstlisting}
\caption{Call stack overhead stats for the Clang and Intel case study 2.}
\label{fig:case2}
\end{figure*}

At the top of the stack, both binaries spend similar amounts of time
in \texttt{start\_thread} from \texttt{libpthread-2.28.so}.
The Clang binary spends $93\%$ of the time in \texttt{\_\_kmp\_invoke\_microtask}
from \texttt{libomp.so}, and the 
Intel spends similar amounts of time, $89\%$, in \texttt{\_\_kmp\_launch\_worker} from
\texttt{libiomp5.so}. From this, we infer that in both cases, the test program
makes the respective OpenMP runtime systems consume overhead time launching and invoking
tasks.

We now look at the performance counters, shown in Table~\ref{table:case2_counters}.
The clang binary incurs much higher counters in many categories,
including much higher number of context switches, page faults, branches and
instructions.
These findings correlate with the source code of the test. The generated code
includes an OpenMP parallel region inside a for loop (a serial loop). Since
the OpenMP implementation spends a lot of time creating tasks, this may explain the high
overheads in task launching in Clang. More analysis, however, is needed
to understand why the binaries coming from the Intel and GCC implementations
are much better at managing OpenMP resources for this test than for the Clang
implementation.


\subsection{Case Study 3: Intel Binary Hangs}
In this case, we analyze an Intel binary that hangs.\footnote{
Refer to file \texttt{quartz1247\_532344/\_tests/\_group\_3/\_test\_3.cpp}}
The binaries from Clang and GCC, however, terminate quickly in a few milliseconds.
We let the Intel binary run for at least 3 minutes, after which we
stop it by sending a \texttt{SIGINT} (CTRL-C) signal to the program.
We run the Intel binary in the \texttt{gdb} debugger, let it run
for 3 minutes, and stop it again.

\begin{figure*} 
    \begin{lstlisting}[
        xleftmargin=0.05\textwidth,
        linewidth=0.98\textwidth, 
        basicstyle=\scriptsize\ttfamily,
        numbers=none
        ] 
^C
Thread 1 "quartz1247_5323" received signal SIGINT, Interrupt.
...
(gdb) bt
#0  0x000015555443a9a8 in __kmp_wait_4 (...) at ../../src/kmp_dispatch.cpp:3118
#1  0x000015555446b49f in _INTERNAL77814fad::__kmp_acquire_queuing_lock_timed_template<false> (...) at ../../src/kmp_lock.cpp:1208
#2  __kmp_acquire_queuing_lock (lck=0x1, gtid=0) at ../../src/kmp_lock.cpp:1254
#3  0x000015555443085d in __kmpc_critical_with_hint (...) at ../../src/kmp_csupport.cpp:1610
#4  0x0000000000402c04 in .omp_outlined._debug__ (...) at quartz1247_532344-_tests-_group_3_test_3.cpp:103
#5  .omp_outlined.(void) const (...) at quartz1247_532344-_tests-_group_3-_test_3.cpp:36
...
    \end{lstlisting}
\caption{GDB backtrace for Thread 1 for Case study 3.}
\label{fig:case3_bt}
\end{figure*}

\begin{figure}
\centering
\includegraphics[width=2.8in]{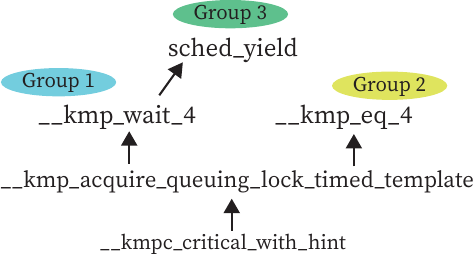}
\caption{State of each thread in Case study 3; the 32 threads
are grouped in three states.}
\label{fig:hang_stack}
\end{figure}

We gather the location of each thread using \texttt{gdb};
there are 32 threads running in total (the number of cores in 
the system). The state of the threads can be grouped into three,
as shown in Figure~\ref{fig:hang_stack}.
All threads are stuck in the function
\texttt{\_\_kmpc\_critical\_with\_hint}, which then calls
\texttt{\_\_kmp\_acquire\_queuing\_lock...}---one group of threads
is in \texttt{\_\_kmp\_wait\_4} (group 1). Another group is in
\texttt{\_\_kmp\_eq\_4} (group 2). A third group is in
\texttt{sched\_yield}, which is called by \texttt{\_\_kmp\_wait\_4}.


Looking at the source code of the test reveals that there is a
critical section in an OpenMP parallel regions, which correlates
with the fact that all threads are stuck in the \texttt{\_\_kmpc\_critical\_with\_hint}
call. Based on this, we hypothesize that the correctness issue
can be caused by either (a) deadlock situation in 
the critical region, or (b) some performance inefficiency in the critical region in terms of 
waiting, spinning, and acquiring locks for the critical region,
which causes the parallel region to not make progress.
Here also more debugging is needed to determine the root cause.


\begin{mdframed}[backgroundcolor=light-gray] \textbf{Answer to Q2:}
Based on the presented case studies, some of the test programs can 
expose possible correctness bugs or performance issues in the tested
OpenMP implementations. Some of the issues involve resource contention,
different overheads for task launching, and possible deadlocks.
\end{mdframed}

\section{Related Work}

Random testing~\cite{duran1984evaluation, hamlet2002} has been used as a 
black-box testing method to perform tests under randomly-generated inputs.
Randomized differential testing~\cite{mckeeman1998differential, groce2007randomized} has 
been used in previous work to detect bugs in compilers. 
A notable instance is Csmith~\cite{csmith}, which detects compiler 
bugs in C compilers. Csmith has found hundreds of bugs in C 
compilers when compiling the programs it generates.
It has also been used as the basis of mutation-based systems, 
where Csmith’s output was modified using other tools to provoke
compiler bugs~\cite{le2014compiler}. 
The CLsmith tool derived from Csmith has been used to find many bugs in OpenCL compilers~\cite{lidbury2015many}. The JTT~\cite{zhao2009automated} program generator is designed to directly test the efficiency and the logic of compiler optimizations.
Laguna presented the \texttt{Varity} framework~\cite{varity}, which generates random
floating-point programs and checks for numerical inconsistencies between
CPUs and GPUs. The original framework, however, did not support generating OpenMP
programs; we have extended \texttt{Varity} to support OpenMP program generation and catching
performance and correctness outliers in this paper.

Several efforts develop custom curated parallel 
benchmarks to study the performance and the scaling
of parallel algorithm implementation. Some of them include the NAS benchmarks~\cite{jin1999openmp},
the PARSEC benchmark suite \cite{bienia2009parsec, de2017bringing},  SPEC~\cite{limaye2018workload, brunst2022first} and the Barchelona OpenMP Task Suite (BOTS)~\cite{5361951} that implements tasking benchmarks.
Task overheads ~\cite{lagrone2011set}. These works provide well tested implementations of common parallel algorithms and can be used as examples for developers. Further, the same benchmarks can be used by various parallel runtimes and compilers to test the efficiency of the respective implementations. In contrast to our work,
these benchmarks require manual and tedious effort to be implemented and maintained. Further, these works use
the most common constructs provided by the parallel programming model. Our work, since it uses random program generation does not require manual effort and can create thousands of tests automatically by traversing the semantically correct grammar and can be used to test performance and correctness of parallel OpenMP programs. 

There are several works comparing the behavior of different compilers. 
The authors in~\cite{machado2017comparing, sr2019battle} compare the
efficiency of different compilers---in terms of execution time 
of the generated executable and the size of  binaries---using 
carefully curated benchmarks or by comparing specific optimizations between compilers, such as loop vectorization~\cite{jubb2014loop}
While in~\cite{sun2016toward} the authors compare the rate
of fixing bug-fixes between GCC and LLVM concluding that 
random program generation can be an efficient approach 
for both correctness test coverage and performance. 
The foundation of these works lies on comparing quantities 
of interest across different tools, our work relies on the same foundation. 
However, in contrast to these works our work generates the 
tests automatically without requiring manual effort.

Besides the implementation of the respective compiler and parallel runtime library there are several works
studying the performance variability induced by the system software (Operating System Jitter) when executing OpenMP parallel programs~\cite{barranco2016analysis, cui2023analysis, mazouz2011analysing}. These works use manually curated parallel OpenMP programs to test the efficiency of the system providing 
insights to procuring hardware and developing system software techniques. 
Recent work finds performance optimization opportunities in OpenMP programs via mutation testing~\cite{DBLP:conf/ppopp/MiaoLGPR24,DBLP:journals/pc/MiaoLGPR24}.
By contrast, our work generates random programs and can be informative to OpenMP runtime developers. 


\section{Conclusion}

Testing implementations of OpenMP is crucial to ensure they
meet required specifications, are defect-free, and are 
ready for production use.
We present an approach to test OpenMP implementations via
random program generation and differential testing.
Our approach generates thousands of program tests for
different OpenMP implementations in a given system, and identifies outliers that could indicate performance or correctness bugs
in OpenMP implementations.
When we evaluate our method with three OpenMP implementations 
(Intel, Clang, and GCC), we identified more than a hundred performance
outlier test cases, and about four correctness outlier cases.
The paper presents several case studies that give 
more details about the possible root cause of these cases.
Future work could involve extending the approach
to support accelerators. This extension would involve
identifying interesting OpenMP accelerator directives,
encoding them in \texttt{Varity}, and extending the grammar to
reflect the allowed programs.

{
\def\bibfont{\footnotesize} 
\bibliographystyle{ieeetr}
\bibliography{biblio}
}

\end{document}